\documentclass[twocolumn,showpacs,preprintnumbers,amsmath,amssymb,floatfix,prd]{revtex4}

\usepackage{citesort}
\usepackage{epsfig}

\begin{document}


\title{Global Analysis of Fragmentation Functions for Protons and Charged Hadrons}
\author{Daniel de Florian}\email{deflo@df.uba.ar}
\author{Rodolfo Sassot}\email{sassot@df.uba.ar}
\affiliation{Departamento de Fisica, Universidad de Buenos Aires, Ciudad Universitaria, Pab.\ 1 (1428)
Buenos Aires, Argentina}
\author{Marco Stratmann}\email{marco@ribf.riken.jp}
\affiliation{Radiation Laboratory, RIKEN, 2-1 Hirosawa, Wako, Saitama 351-0198, Japan}

\begin{abstract}
We present new sets of fragmentation functions for protons and inclusive
charged hadrons obtained in combined NLO QCD analyses of single-inclusive
hadron production in electron-positron annihilation, proton-proton collisions,
and deep-inelastic lepton-proton scattering. These analyses complement
previous results for pion and kaon fragmentation functions with charge and flavor
discrimination. The Lagrange multiplier technique is used to assess the uncertainties
in the extraction of the new sets of fragmentation functions.
\end{abstract}


\maketitle

\section{Introduction}

Single-inclusive hadron production is the most appropriate
and powerful benchmark for challenging altogether our understanding of
the partonic structure of nucleons, the dynamics of hard
Quantum Chromodynamics (QCD) interactions, the validity of QCD
factorization, and the way in which quarks and gluons produce
the detected final-state hadrons.
The last few years have witnessed a
remarkable improvement in both precision and variety for this kind of measurements,
which are expected to continue yielding crucial pieces of information in the future.

In a recent article \cite{deFlorian:2007aj}, we have demonstrated the
feasibility of performing a next-to-leading order (NLO) combined QCD
analysis of single-inclusive pion and kaon production data, coming from
electron-positron annihilation, semi-inclusive deep-inelastic lepton-nucleon scattering (SIDIS),
and proton-proton collisions, collected over a wide kinematic range. The
analysis not only allowed a very good {\em global} description of all these processes,
but provided also accurate parametrizations for the parton-to-pion and parton-to-kaon fragmentation
functions, which encode the details of the non-perturbative hadronization process relevant for the
perturbative QCD (pQCD) framework based on the factorization theorem.

In the following, we extend the analysis of Ref.~\cite{deFlorian:2007aj}
first to the case of single-inclusive proton and anti-proton production, again
including not only the electron-positron annihilation data
\cite{ref:tpcdata,ref:slddata,ref:alephdata,ref:delphidata,ref:opaleta}
used in all previous fits \cite{ref:kkp,ref:akk,ref:hirai}, but, for the first time,
also the very precise data recently obtained in proton-proton collisions
at RHIC \cite{ref:starproton}, where final state protons and anti-protons are discriminated.

Next, after having obtained reliable fragmentation functions for the three lightest and
most copiously produced charged hadron species, pions and kaons from Ref.~\cite{deFlorian:2007aj}
and (anti-)protons, we analyze inclusive, i.e., {\em unidentified}, charged hadron yields obtained
in electron-positron annihilation
\cite{ref:tpcdata,ref:slddata,ref:alephdata,ref:delphidata,ref:tassodata,ref:opal,ref:opall,ref:delphil},
proton-(anti-)proton collisions \cite{ref:tev,ref:ua1,ref:ua2}, and
SIDIS \cite{ref:emc}.
Here we are aiming at an extraction of the contribution from the
``residual'' charged hadrons other than pions, kaons, and
(anti-)protons, to the inclusive charged hadron fragmentation functions.

Only from a global QCD analysis we can obtain individual quark and anti-quark fragmentation functions for unidentified
positively or negatively charged hadrons without having to make assumptions on the relation between
favored (valence-type) and unfavored (sea-type) contributions.
Such assumptions have been shown to be often not adequate for reproducing presently available
single-inclusive hadron production yields beyond those obtained in
electron-positron annihilation \cite{deFlorian:2007aj}.
In addition, data from proton-(anti-)proton collisions are crucial for reducing the uncertainties in
the gluon-to-hadron fragmentation function \cite{deFlorian:2007aj} since scaling violations
of $e^+e^-$ data are not much of a constraint due to the lack of precision data at more than
one value of the center-of-mass system (c.m.s.) energy $\sqrt{s}$.
Both resulting new sets of NLO fragmentation functions reproduce with remarkable accuracy the data available
and hence complement consistently our previous studies for pions and kaons.

Previously available sets of inclusive charged hadron fragmentation functions
are all based on fits to $e^+e^-$ data only and either sum up
the results obtained for pions, kaons, and (anti-)protons, ignoring
possible contributions from heavier charged hadrons \cite{ref:kkp,ref:akk}
or the residual hadron contribution already includes protons and anti-protons
\cite{ref:kretzer}. The analysis of Ref.~\cite{Bourhis:2000gs} provides only fragmentation
functions for unidentified charged hadrons, again based on $e^+e^-$ data only, however,
an attempt was made to further constrain the gluon fragmentation function by
comparing to one of the many data sets from proton-(anti-)proton collisions.

In the next section, we briefly highlight the main features of the present
analysis, specifically discussing our choice of parametrizations for the proton
and the residual charged hadron fragmentation functions
and emphasizing the charge conjugation and flavor symmetry assumptions we still have to make.
In Section III we present our results for both NLO fits, and assess the uncertainties
involved with the help of the Lagrange multiplier technique.
In Sec. IV we summarize our main results.
For completeness, the Appendix contains the results of the global analyses performed
at leading order (LO) accuracy, which compares significantly less favorable
to data than our NLO fits.

\section{\label{sec:outline} Outline}
%
In the present analysis we work in the well established framework of pQCD
for single-inclusive hadron production processes at NLO accuracy,
thoroughly discussed and implemented in Ref.~\cite{deFlorian:2007aj}.
We make an extensive use of the Mellin transform technique \cite{ref:mellin}
developed for a fast computation of the exact NLO cross sections in each
step of a global $\chi^2$ minimization procedure.
We refer the reader to \cite{deFlorian:2007aj}
and references therein for further technical details on the general
QCD framework and the assessment of uncertainties in global fits
with the help of Lagrange multipliers.
The analysis proceeds in two stages, the first dedicated to the extraction of the
parton-to-(anti-)proton fragmentation functions, followed by the determination of
inclusive charged hadron fragmentation functions in the second.

\subsection{\label{sec:proton} Proton fragmentation functions}
In the first step of our global analysis we aim to determine individual
fragmentation functions $D_i^H$ for quarks and anti-quarks of all flavors $i$ as well
as for gluons into either protons ($H=p$) or anti-protons ($H=\bar{p}$).
At variance with previous analyses \cite{ref:kkp,ref:akk,ref:hirai}, the use of data that
discriminate between $p$ and $\bar{p}$ in the final-state, allows us to extract
fragmentation functions for either particle. As explained in Ref.~\cite{deFlorian:2007aj},
in order to have the flexibility required by charge separated distributions
and to accommodate the additional data, we adopt a somewhat more versatile functional
form for the input distributions at scale $\mu_0$ than in \cite{ref:kkp,ref:akk,ref:hirai}
\begin{eqnarray}
\label{eq:ff-input}
 D_i^H(z,\mu_0) =   \hspace{6cm}  \nonumber \\
 \hspace{1cm} \frac{N_i z^{\alpha_i}(1-z)^{\beta_i} [1+\gamma_i (1-z)^{\delta_i}] }
{B[2+\alpha_i,\beta_i+1]+\gamma_i B[2+\alpha_i,\beta_i+\delta_i+1]},
\end{eqnarray}
where $B[a,b]$ represents the Euler beta function and $N_i$ is
normalized such to represent the contribution of $D_i^H$ to the second moment
$\int_0^1 dz z D_i^H(z,\mu_0)$ entering the energy-momentum sum rule.
$z$ denotes the fraction of the fragmenting parton's energy taken by the produced
hadron $H$.
As in \cite{deFlorian:2007aj}, the initial scale $\mu_0$ in Eq.~(\ref{eq:ff-input})
for the $Q^2$-evolution is taken to be $\mu_0=1\,\mathrm{GeV}$ for the light $u,\,d,\,s$ partons and the 
quark masses for the heavier ones.

The parameters describing the fragmentation functions $D_i^H(z,\mu_0)$
in Eq.~(\ref{eq:ff-input}) are determined by a standard $\chi^2$ minimization
for $K$ data points, where
\begin{equation}
\label{eq:chi2}
\chi^2=\sum_{j=1}^K \frac{(T_j-E_j)^2}{\delta E_j^2},
\end{equation}
$E_j$ is the measured value of a given observable,
$\delta E_j$ the error associated with this measurement, and
$T_j$ is the corresponding theoretical estimate for a
given set of parameters in (\ref{eq:ff-input}).
Since the full error correlation matrices are not available for
most of the data entering the global analysis, we
take the statistical and systematical errors in quadrature
in $\delta E_j$.

In order to reduce the number of parameters in (\ref{eq:ff-input}) to those
that can be effectively constrained by the data, we impose, as usual, certain
relations among the individual fragmentation functions.
We have checked in each case that relaxing these assumptions indeed does not
significantly improve the total $\chi^2$ of the fit (\ref{eq:chi2}) to warrant any additional parameters.
In detail, for the fragmentation of light quark flavors to a proton
we assume the same shape for up and down quarks and for up and down
anti-quarks with the same $z$-independent normalization ratios $N$, i.e.,
\begin{equation}
\label{eq:iso}
D_{{u}}^{p}= N D_{d}^{p}\,\,\,\,\,\,\,\, \mbox{and}\,\,\,\,\,\,\,\,  D_{\bar{u}}^{p}= N D_{\bar{d}}^{p}.
\end{equation}
The relation between quark and anti-quark fragmentation functions for $u$ and $d$ flavors
is determined by the global fit through
\begin{equation}
\label{eq:charge}
2\,D_{\bar{q}}^{p}= (1-z)^{\beta} D_{q+\bar{q}}^{p},
\end{equation}
with $\beta$ constrained to be positive.
For strange quarks it is assumed that
\begin{equation}
\label{eq:sea_break}
D_s^{p}=D_{\bar{s}}^{p} =N^{\prime} D_{\bar{u}}^{p},
\end{equation}
with the SU(3)-breaking parameter $N^{\prime}$ independent of $z$.
For charm, bottom, and gluon-to-proton fragmentation we find no improvement in the total $\chi^2$
for $\gamma_i\neq 0$ in (\ref{eq:ff-input}), hence we set $\gamma_i=0$.

To obtain the corresponding fragmentation functions $D^{\bar{p}}_i$ for anti-protons
we assume charge conjugation symmetry, i.e.,
\begin{equation}
D_{{q}}^{p}=D_{\bar{q}}^{\bar{p}},
\end{equation}
leaving in total 17 free parameters to be determined by the global fit.

\subsection{\label{sec:chghad} Charged hadron fragmentation functions}
The fragmentation functions $D_i^H$ for
unidentified positively charged hadrons ($H=h^+$) are defined by
\begin{equation}
\label{eq:residual}
D_i^{h^+}=D_{i}^{\pi^+}+D_{i}^{K^+}+D_{i}^{p}+D_i^{res^+},
\end{equation}
where $D_{i}^{\pi^+}$ and $D_{i}^{K^+}$ were already determined in Ref.~\cite{deFlorian:2007aj}.
$D_i^{res^+}$ denotes the residual contribution of positively charged hadrons other than
pions, kaons, and protons to the inclusive sum $D_i^{h^+}$. A definition analogous to
Eq.~(\ref{eq:residual}) is used for the fragmentation $D_i^{h^-}$
into negatively charged hadrons.

Since pions are much more copiously produced than heavier kaons and protons are even less abundant, it
is natural to expect that Eq.~(\ref{eq:residual}) is strongly dominated
by $D_{i}^{\pi^+}+D_{i}^{K^+}+D_{i}^{p}$ leaving $D_i^{res^+}$ to be small.
Nevertheless it is an important consistency check to extract a small but
non-vanishing $D_i^{res^+}$ (and $D_i^{res^-}$) from data to actually confirm this hierarchy.
If the global fit would require, for instance, a large or even a negative residual contribution,
the usefulness of the previously extracted fragmentation functions for light hadrons
would be in jeopardy.

For $D_i^{res^+}$ it turns out that the data are most economically described by
assuming full SU(3) flavor symmetry for both quarks and anti-quarks
\begin{equation}
D_{{u}}^{res^+}=D_{{d}}^{res^+}=D_{{s}}^{res^+}
\end{equation}
and
\begin{equation}
D_{\bar{u}}^{res^+}=D_{\bar{d}}^{res^+}=D_{\bar{s}}^{res^+},
\end{equation}
respectively, in the ansatz (\ref{eq:ff-input}). Again,
\begin{equation}
\label{eq:charge2}
2\,D_{\bar{q}}^{res^+}= (1-z)^{\beta^{\prime}} D_{q+\bar{q}}^{res^+}
\end{equation}
and for $D_{c,b}^{res^+}$ we set $\gamma_{c,b}=0$ in (\ref{eq:ff-input}),
however, $D_{g}^{res^+}$ has a preference for $\gamma_g\neq 0$.
For  $D_i^{res^+}$ we assume charge conjugation symmetry, i.e.,
\begin{equation}
D_{{q}}^{res+}=D_{\bar{q}}^{res-},
\end{equation}
leaving in total 18 free parameters to be fitted here.

We note that the $Q^2$-dependence for all fragmentation functions is computed with the
appropriate NLO evolution equations as explained in detail in Ref.~\cite{deFlorian:2007aj}.
Uncertainties in the extraction of the parameters in Eq.~(\ref{eq:ff-input}) will
be assessed with the help of the Lagrange multiplier technique as described again
in \cite{deFlorian:2007aj} and references therein.

\section{\label{sec:results} Results of the global analyses}
%
In this section we discuss in detail the results of our global
analyses of fragmentation functions \footnote{A {\tt Fortran} package 
containing our LO and NLO sets of 
fragmentation functions can be obtained upon request from the authors.} 
for protons and residual charged hadrons
as outlined above.
We present the parameters of the optimum fits
describing the NLO $D_i^{p,res^+}$ at the input scale $\mu_0$ and
compare to all data sets used in the analyses, including
individual $\chi^2$ values.
Detailed comparisons are made with previous fits
based exclusively on the available electron-positron annihilation data
in Refs.~\cite{ref:akk}, \cite{ref:hirai}, and \cite{ref:kretzer},
in the following  labeled as AKK, HKNS, and KRE, respectively.

\subsection{NLO analysis of proton fragmentation functions}
%
Three different processes allow the extraction of the (anti-)proton
fragmentation functions in our global analysis.
First of all, we have the ``standard'' electron-positron
annihilation data \cite{ref:tpcdata,ref:slddata,ref:alephdata,ref:delphidata}
customarily included also in most of the previous analyses
\cite{ref:kkp,ref:akk,ref:hirai}.
A characteristic feature of
$e^+e^-$ annihilation data in general is that they only provide information on a
certain hadron species summed over the charges, i.e., in this case
on the sum of protons and anti-protons. Besides the
fully inclusive measurements, SLD \cite{ref:slddata} and DELPHI \cite{ref:delphidata}
also give results for ``flavor enriched'' cross sections, distinguishing
between the sum of light $u,\, d,\, s$ quarks, charm, and bottom events. The quark
flavor is determined from Monte-Carlo simulations and refers to the primary $q\bar{q}$ pair
created by the intermediate $Z$-boson or photon.
Such results have to be taken with a grain of salt as they cannot be unambiguously
interpreted and calculated in pQCD. Nevertheless, in our analysis of pion and
kaon data \cite{deFlorian:2007aj} we always found good agreement between data
and theory.

\begin{figure*}[t!]
\begin{center}
\vspace*{-0.6cm}
\epsfig{figure=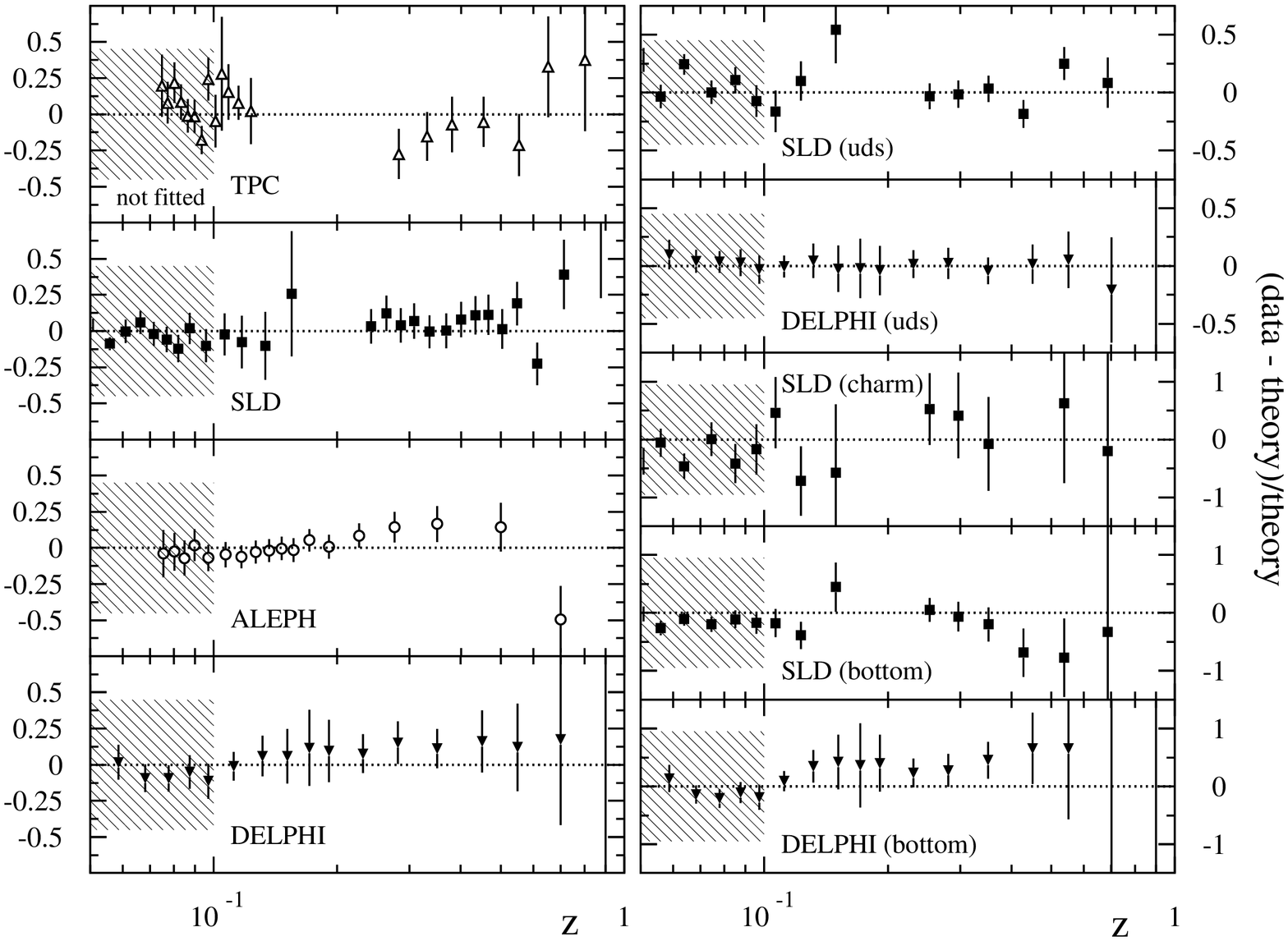,width=0.75\textwidth}
\end{center}
\vspace*{-0.7cm}
\caption{``(data-theory)/theory'' comparison of our NLO results for the
electron-positron annihilation cross section into protons and anti-protons
with the data sets used in the fit, see also Tab.~\ref{tab:expprotontab}.
\label{fig:protoninclusive}}
\vspace*{-0.5cm}
%
\begin{center}
\epsfig{figure=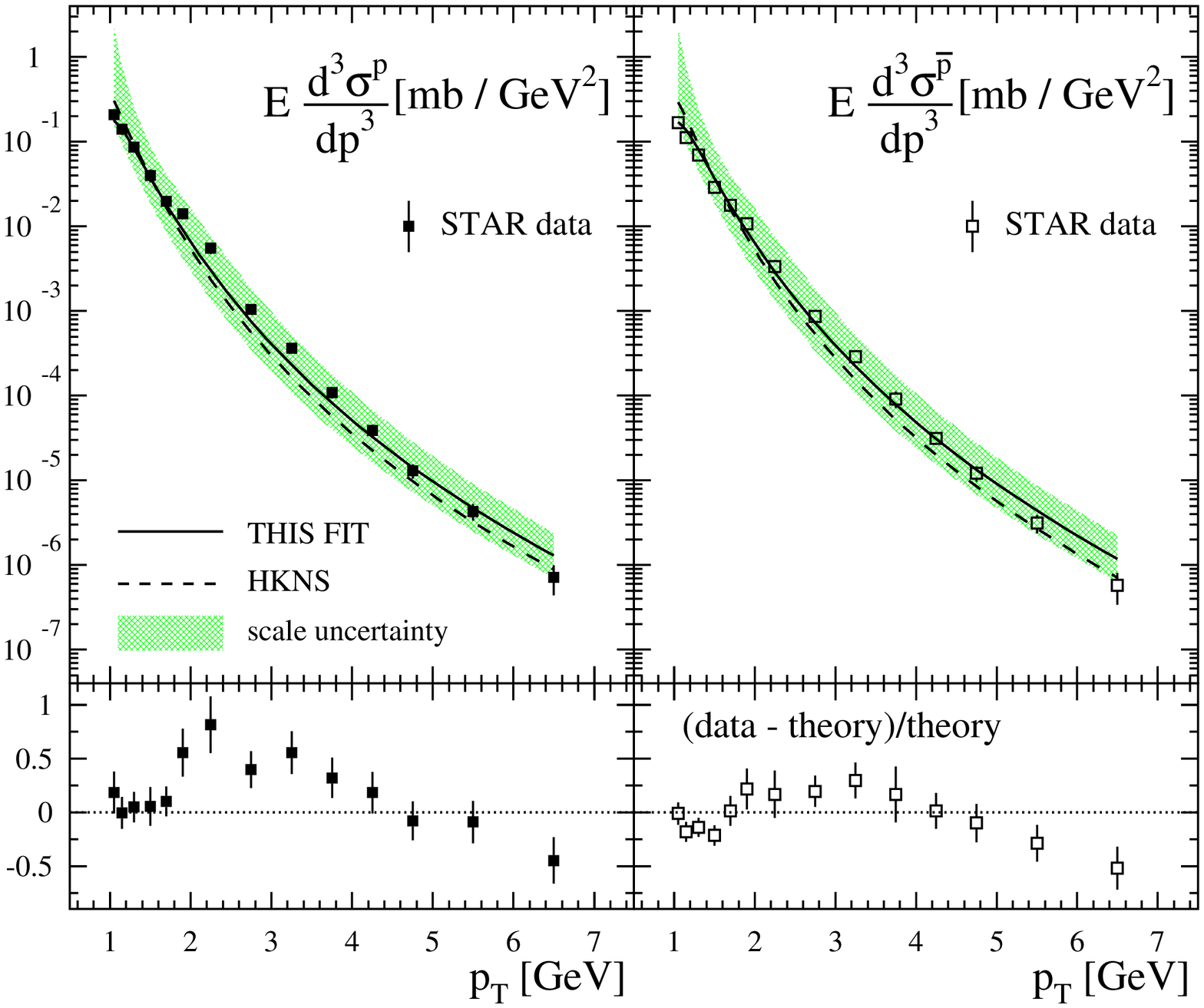,width=0.75\textwidth}
\end{center}
\vspace*{-0.7cm}
\caption{Upper panels: comparison of our NLO results (solid lines) for
single-inclusive proton (anti-proton) production $pp\rightarrow p(\bar{p}) X$
at $\sqrt{s}=200\,\mathrm{GeV}$ with STAR data \cite{ref:starproton} at mid-rapidity $|y|<0.5$ using
$\mu_f=\mu_r=p_T$.
Also shown are the results obtained with the HKNS \cite{ref:hirai} parametrization.
The shaded bands indicate theoretical uncertainties when all
scales are varied in the range $p_T/2\le \mu_f=\mu_r \le 2 p_T$.
Lower panels: ``(data-theory)/theory'' for our NLO results.
\label{fig:star-proton}}
\vspace*{-0.5cm}
\end{figure*}

In Fig.~\ref{fig:protoninclusive} we provide a detailed comparison
in terms of ``(data-theory)/theory'' between all the $e^+e^-$ data sets included
in the analysis and our NLO results.
The best agreement is found for ALEPH and DELPHI inclusive measurements,
although SLD and TPC, as well as flavor tagged data sets, also agree well
with the fit within the fairly large experimental uncertainties.
The fluctuations in the data are significantly larger than in the case
of pions, but roughly comparable to those found for kaons, see Ref.~\cite{deFlorian:2007aj}.
Other recent fits \cite{ref:akk,ref:hirai} all reproduce the data shown in
Fig.~\ref{fig:protoninclusive} equally well within the experimental uncertainties.
Since the current $e^+e^-$ data basically only constrain the total singlet fragmentation
function $D_{\Sigma}^{p+\bar{p}}$ at scale $M_Z$, all sets differ considerably
in their individual flavor content as we shall see below.

We wish to recall that the range of applicability for fragmentation functions is
severely limited to medium-to-large values of $z$, see, e.g., Ref.~\cite{deFlorian:2007aj}.
In order to avoid the potentially problematic low-$z$ region, we exclude from the fit
all data with energy fractions lower than $z_{\min}=0.1$. The extrapolation of our fit, however,
reproduces the trend of the data reasonably well also below $z_{\min}$ as indicated
in Fig.~\ref{fig:protoninclusive}.
%

The second key ingredient in our global analysis are the single-inclusive hadron production
data from proton-proton collisions at BNL-RHIC
taken by STAR \cite{ref:starproton} at mid-rapidity $|y|<0.5$ and shown in Fig.~\ref{fig:star-proton}.
These data discriminate final-state protons from anti-protons and hence, in principle,
allow to separate quark-to-proton and anti-quark-to-proton fragmentation functions in the fit.
However, at the presently accessible range of transverse momenta $p_T$ and at mid-rapidities
the production of single-inclusive hadrons is mainly driven by gluon-induced processes and fragmentation,
turning these data into the best constraint on the gluon fragmentation function $D_g^p$
at large values of $z$ currently available.
Figure \ref{fig:star-proton} shows the results of the fit compared to the data.
The theoretical uncertainties related to a particular choice of the factorization and renormalization
scales, $\mu_f$ and $\mu_r$, respectively, in the computation of the NLO cross section,
indicated by the shaded bands, are non-negligible.
This has been taken into account in the $\chi^2$ minimization procedure as a
conservative additional 5\% relative uncertainty.

\begin{figure}[t!]
\begin{center}
\vspace*{-0.6cm}
\epsfig{figure=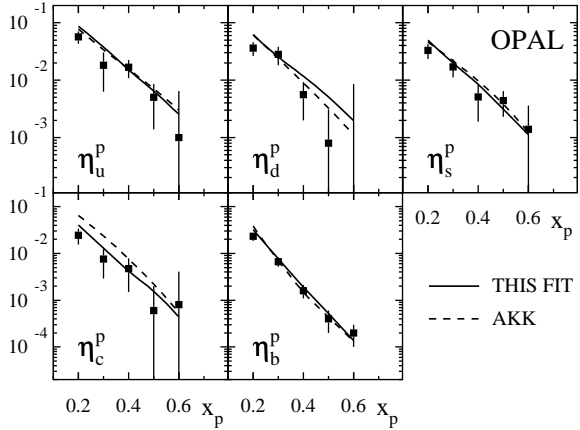,width=0.5\textwidth}
\end{center}
\vspace*{-0.7cm}
\caption{Comparison of the OPAL ``tagging probabilities''
\cite{ref:opaleta} for protons plus anti-protons, as a
function of the minimum $x_p$,  with our NLO results (solid lines).
Also shown are the results obtained with the
AKK \cite{ref:akk} parameterization (dashed lines).
\label{fig:opal-proton-eta}}
\vspace*{-0.5cm}
\end{figure}
We have checked that by excluding data
at low $p_T$ values, $p_T\lesssim 3\,\mathrm{GeV}$, the results of the global fit remain essentially
unchanged. The fact that the $p_T$ spectrum is well reproduced down to $1\,\mathrm{GeV}$,
the mass of the proton, may turn out to be accidental, but this can only be verified once data
at higher $p_T$ become available to map out the $p_T$ slope more precisely.
We note that the BRAHMS data \cite{ref:brahmsdata} at very forward rapidities $y\simeq 3$
and $p_T\lesssim 4\,\mathrm{GeV}$ have not been included in our global analysis.
These data show a very pronounced difference between the
$p$ and $\bar{p}$ yields which cannot be understood.
The origin of the excess of protons over anti-protons by a factor of about 10, not observed 
by STAR at less extreme kinematics, remains an open question \cite{ref:brahmsdata}.

Finally, a third source of information on the parton-to-proton fragmentation
is provided by the OPAL ``tagging probabilities'' $\eta_i^p$
\cite{ref:opaleta}, also included in the AKK fit \cite{ref:akk} but not
in \cite{ref:hirai}, which are sensitive to the
flavor of $q+\bar{q}$ fragmentation functions.
The $\eta_i^p(x_p)$ represent ``probabilities'' for a quark flavor $i$ to produce
a ``jet'' containing the (anti-)proton with a momentum fraction $z$ larger than $x_p$.
Figure \ref{fig:opal-proton-eta} compares the results from our and
the AKK fit with the data.
As discussed at length in Ref.~\cite{deFlorian:2007aj} and also mentioned in the
discussion of the flavor tagged SLD and DELPHI data above, any flavor tagged information
is highly model dependent and difficult to interpret within pQCD beyond the LO approximation
involve non trivial theoretical uncertainties.
To take this into account, we assign an up to $10\%$ extra normalization uncertainty to
the $\eta_i^p$ data in the $\chi^2$ minimization. The agreement between
OPAL data and theory is reasonably good.

Tables \ref{tab:nloprotonpara} and \ref{tab:expprotontab}, show the values
obtained for the parameters in Eq.~(\ref{eq:ff-input}) specifying the
optimum fit of proton fragmentation functions $D_i^p(z,\mu_0)$ at
NLO accuracy and summarize the $\chi^2$ values for each
individual set of data included in the global analysis, respectively.
In cases where the normalization uncertainty of the experiment is not
included in the error bars of the data,
we apply a free normalization factor constrained to vary within range quoted
by the experiment. The values of the normalization factors resulting from the fit
are also included in Tab.~\ref{tab:expprotontab}.
%
%
\begin{table}[bht!]
\caption{\label{tab:nloprotonpara}Parameters describing the NLO
fragmentation functions for protons, $D_i^{p}(z,\mu_0)$,
in Eq.~(\ref{eq:ff-input}) at the input scale $\mu_0=1\,\mathrm{GeV}$.
Inputs for the charm and bottom fragmentation functions refer to
$\mu_0=m_c=1.43\,\mathrm{GeV}$ and
 $\mu_0=m_b=4.3\,\mathrm{GeV}$, respectively. }
\begin{ruledtabular}
\begin{tabular}{cccccc}
flavor $i$ &$N_i$ & $\alpha_i$ & $\beta_i$ &$\gamma_i$ &$\delta_i$\\
\hline
$u+\overline{u}$& 0.091&-0.222& 1.414&15.0& 3.29\\
$d+\overline{d}$& 0.058&-0.222& 1.414&15.0& 3.29\\
$\overline{u}$  & 0.034&-0.222& 2.024&15.0& 3.29\\
$\overline{d}$  & 0.022&-0.222& 2.024&15.0& 3.29\\
$s+\overline{s}$& 0.043&-0.222& 2.024&15.0& 3.29\\
$c+\overline{c}$& 0.076&-0.899& 5.920& 0.0& 0.00\\
$b+\overline{b}$& 0.044&-0.034&10.000& 0.0& 0.00\\
$g$& 0.014& 6.000& 1.200& 0.0& 0.00\\
\end{tabular}
\end{ruledtabular}
\vspace*{0.3cm}
\caption{\label{tab:expprotontab}Data used in the NLO global analysis
of proton fragmentation functions, the individual $\chi^2$ values for
each set, the fitted normalizations, and the total $\chi^2$ of the fit.}
\begin{ruledtabular}
\begin{tabular}{lcccc}
experiment& data & rel.\ norm.\ &data points & $\chi^2$ \\
          & type & in fit       & fitted     &         \\\hline
TPC \cite{ref:tpcdata}  & incl.\  &  1.06  & 12 & 7.7 \\
SLD \cite{ref:slddata}  & incl.\  &  0.983 & 18 & 12.0 \\
          & ``$uds$ tag''         &  0.983 & 9  & 10.5 \\
          & ``$c$ tag''           &  0.983 & 9  &  9.8  \\
          & ``$b$ tag''           &  0.983 & 9  &  8.9 \\
ALEPH \cite{ref:alephdata}    & incl.\  & 0.97 & 13 &  11.5 \\
DELPHI \cite{ref:delphidata}  & incl.\  & 1.0  & 12 & 3.9 \\
          & ``$uds$ tag''   &  1.0  & 12 & 0.6 \\
          & ``$b$ tag''     &  1.0  & 12 & 9.1 \\
OPAL \cite{ref:opaleta}
          & ``$u$ tag'' &  1.10  & 5 & 7.6 \\
          & ``$d$ tag'' &  1.10  & 5 & 13.5  \\
          & ``$s$ tag'' &  1.10  & 5 & 5.0 \\
          & ``$c$ tag'' &  1.10  & 5 & 4.9 \\
          & ``$b$ tag'' &  1.10  & 5 & 5.5 \\\hline
STAR \cite{ref:starproton}  & $p$  & 0.95  & 14  & 35.4           \\
                          & $\bar{p}$ & 0.95  & 14  & 26.0           \\
 \hline\hline
{\bf TOTAL:} & & & 159 & 171.9\\
\end{tabular}
\end{ruledtabular}
\end{table}

\begin{figure}[thb!]
\begin{center}
\vspace*{-0.6cm}
\epsfig{figure=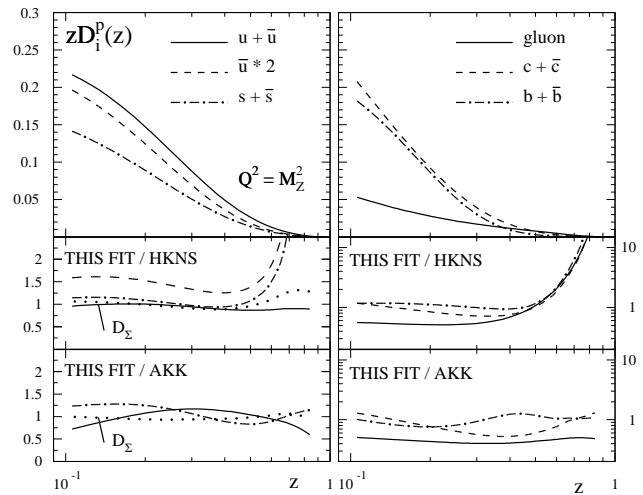,width=0.5\textwidth}
\end{center}
\vspace*{-0.5cm}
\caption{Upper panels: individual fragmentation functions for protons
$zD_i^{p}(z,Q^2)$ at $Q^2=M_Z^2$.
Middle panels: ratios of our fragmentation functions to the ones
of HKNS \cite{ref:hirai}. The dotted line indicates the ratio for singlet
combination of fragmentation functions $zD_{\Sigma}^{p}$.
Lower panels: ratios of our fragmentation functions to the ones
of AKK \cite{ref:akk}.
\label{fig:ff-proton}}
\vspace*{-0.5cm}
\end{figure}
\begin{figure}[h!]
\begin{center}
\vspace*{-.6cm}
\epsfig{figure=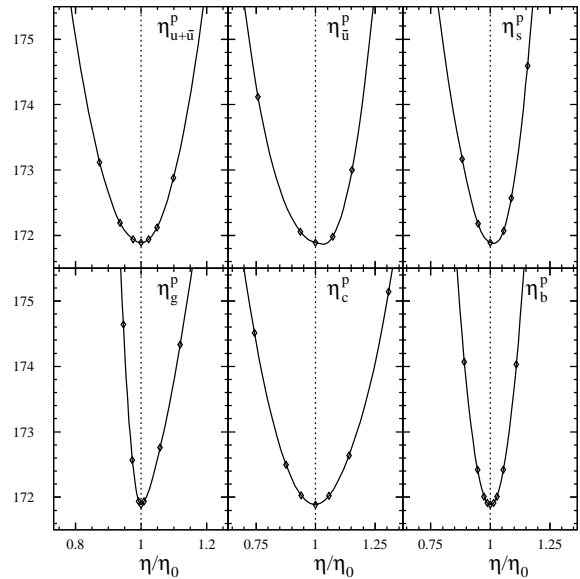,width=0.53\textwidth}
\end{center}
\vspace*{-.7cm}
\caption{Profiles of $\chi^2$ for the NLO proton fragmentation fit
as a function of the truncated second moments
$\eta^{p}_i(x_p=0.2,Q^2=25\,\text{GeV}^2)$ for different flavors.
The moments are
normalized to the value $\eta^{p}_{i\,0}$ they take in the best fit
to data.
\label{fig:proton-uncert}}
\vspace*{-0.5cm}
\end{figure}
Figure \ref{fig:ff-proton} shows the resulting new set of fragmentation functions for
protons for different flavors, evolved to the scale of the mass of the Z-boson,
and in the middle and lower panels, comparisons to the fits of HKNS \cite{ref:hirai}
and AKK \cite{ref:akk}, respectively.
As expected, the present fit agrees well in the singlet combination $D_{\Sigma}^p$
with previous extractions based only on electron-positron annihilation data,
but we find significant differences in the
gluon fragmentation function, constrained in our fit not only by the scale
dependence of the data but also by the STAR data shown in Fig.~\ref{fig:star-proton}.
In addition, there is a sizable difference between our $D_{\bar{u}}^p$ and that of
HKNS which is mainly a consequence of the SU(3) flavor symmetry imposed for
the unfavored fragmentation functions in the latter analysis \cite{ref:hirai}.

As in Ref.~\cite{deFlorian:2007aj}, we make use of the Lagrange multiplier technique
\cite{ref:cteq} in order to give a representative picture of the typical uncertainties characteristic
of the fragmentation functions obtained from the global fit.
In Fig.~\ref{fig:proton-uncert} we show the $\chi^2$-profiles
as a function of the range of variation of the
truncated second moments of the individual fragmentation functions of flavor $i$,
\begin{equation}
\label{eq:truncmom}
\eta^p_i(x_p,Q^2) \equiv \int_{x_p}^1 z D_i^p(z,Q^2) dz,
\end{equation}
for $x_p=0.2$ and $Q=5\, \text{GeV}$, around the values obtained
for them in the best fit to data, $\eta^p_{i\,0}$. Roughly speaking,
the typical uncertainties in the second moments of the proton fragmentation
functions range between 20\% and 25\%, allowing
for a conservative increase $\Delta \chi^2$ of $2\%$ in the total $\chi^2$ of the fit,
except for the gluon and bottom fragmentation functions, where the
uncertainties are closer to $10\%$.  The rather stringent constraint for
bottom comes from the availability of DELPHI and SLD flavor tagged data, while
for charm only SLD data with larger errors are available. Note that neither
charm nor bottom fragmentation play any role in the description of the
STAR data due to the relatively low scales of ${\cal{O}}(p_T)$ involved. For the gluon
fragmentation function $D_g^p$, the STAR data make the difference.
We wish to stress that compared to pions and kaons \cite{deFlorian:2007aj},
the parton-to-proton fragmentation functions $D_i^p$ are much less
constrained at the moment.
This is also reflected in the stronger assumptions, Eqs.~(\ref{eq:iso})-(\ref{eq:sea_break}),
which have to be imposed on the fit in order to be able to determine all parameters.
In particular the lack of (anti-)proton production data from SIDIS
prevent a more reliable separation of favored and unfavored fragmentation functions.
Also the gluon fragmentation is currently mainly determined from a single set of data (STAR). Here,
possible future high precision data from B-factories would open up the possibility
for studies of scaling violations in $e^+e^-$ annihilation, which should help to further constrain
$D_g^p$.

\subsection{NLO analysis of charged hadrons fragmentation functions}
%
\begin{figure*}[th!]
\begin{center}
\vspace*{-0.6cm}
\epsfig{figure=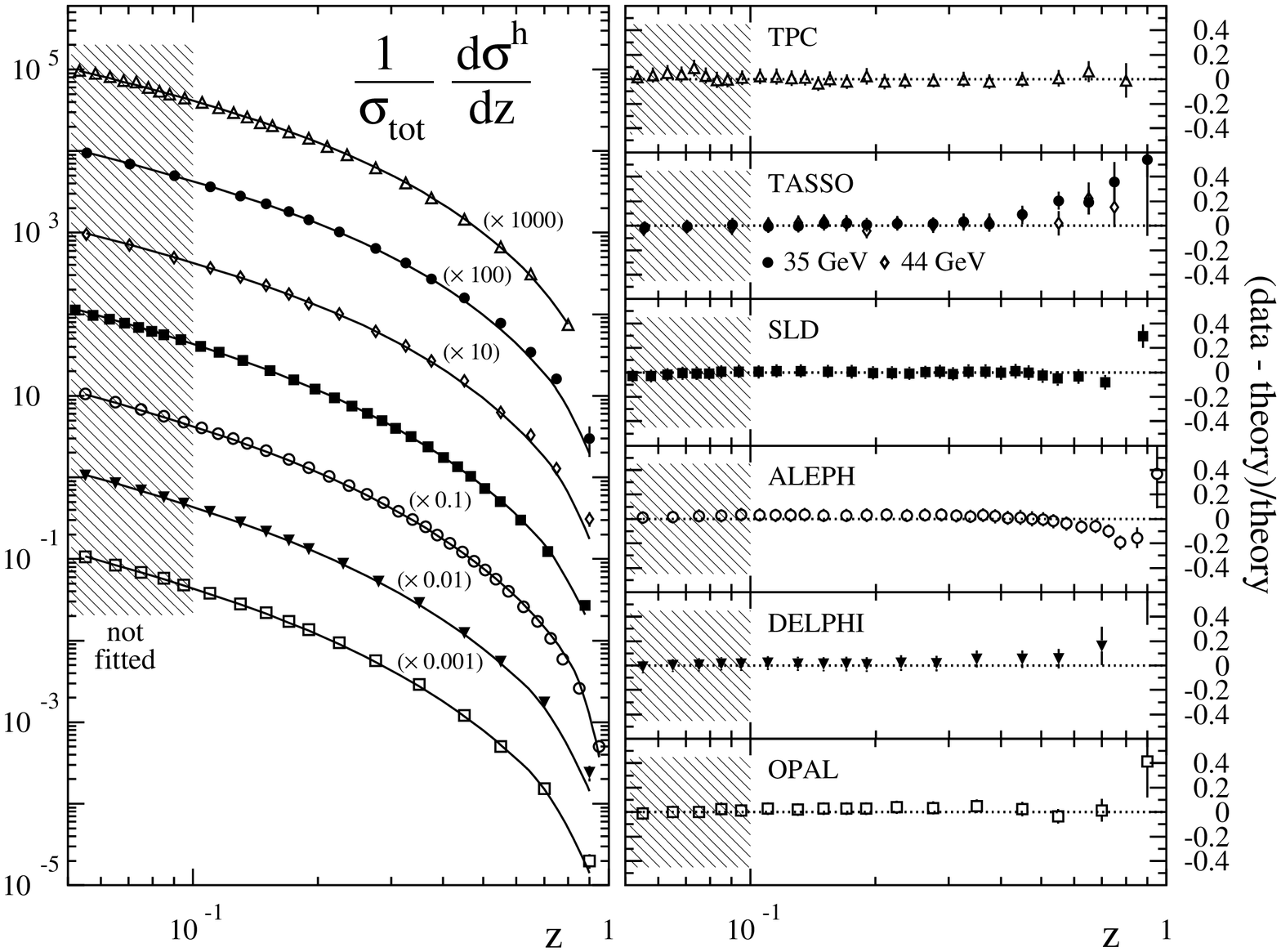,width=0.795\textwidth}
\end{center}
\vspace*{-0.7cm}
\caption{Left-hand side: comparison of our NLO results for the
electron-positron annihilation cross section into inclusive charged hadrons
with the data sets used in the fit, see Tab.\ \ref{tab:exprestab}.
Right-hand side: (data-theory)/theory for our NLO results for each of the data sets.
\label{fig:hadroninclusive}}
%
\begin{center}
\vspace*{-0.6cm}
\epsfig{figure=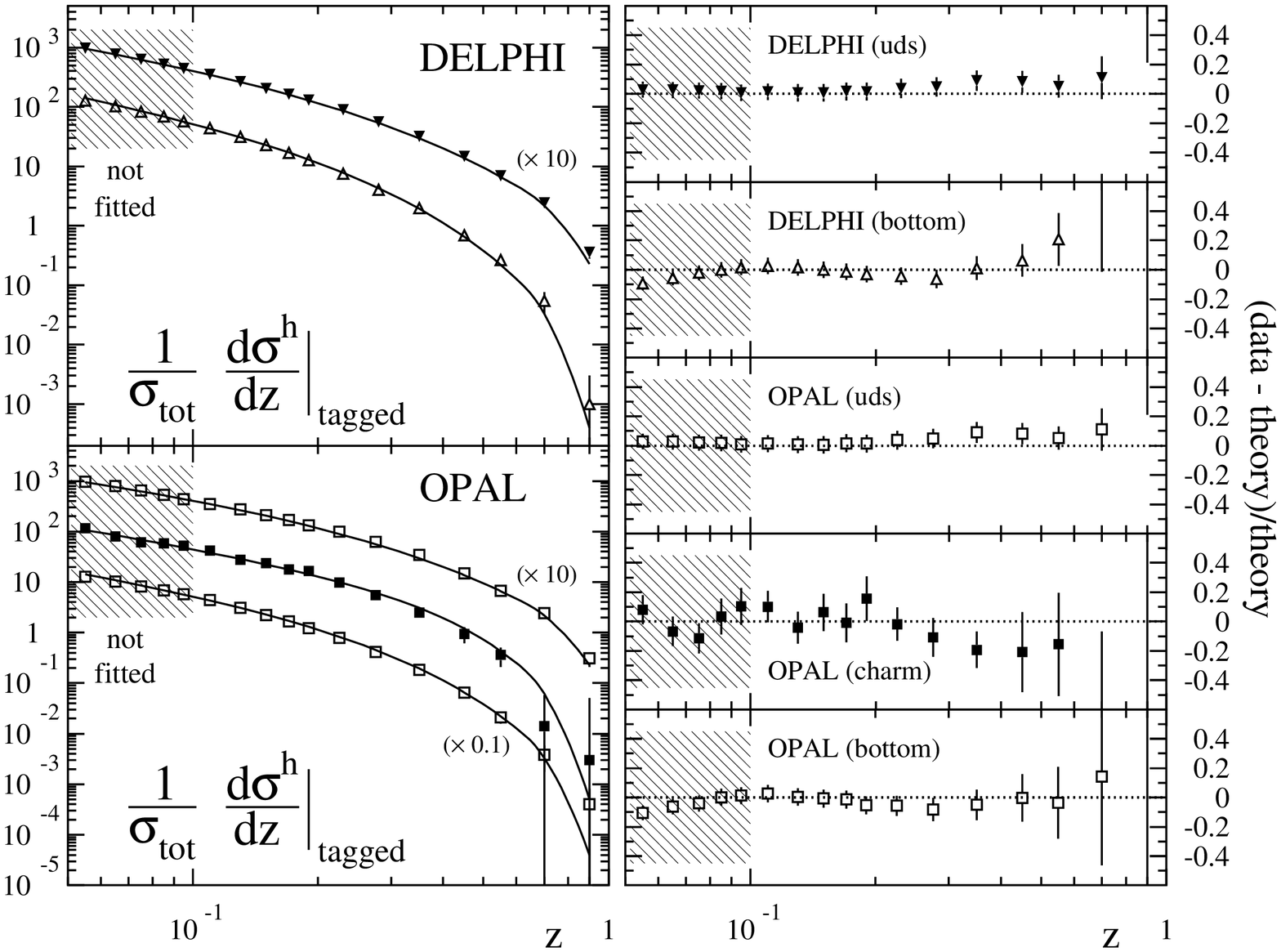,width=0.795\textwidth}
\end{center}
\vspace*{-0.7cm}
\caption{Same as in Fig.\ \ref{fig:hadroninclusive} but now for light (``uds'') and heavy quark (``c,b'') tagged
cross sections.
\label{fig:hadron-tagged}}
\vspace*{-0.5cm}
\end{figure*}
As outlined in Sec.~\ref{sec:chghad},
rather than extracting fragmentation functions for inclusive
charged hadrons from scratch, we take advantage of the knowledge already
acquired on pion, kaon, and proton fragmentation functions.
In the following, we isolate and determine the contributions coming
just from residual charged hadrons $D_i^{res^{\pm}}$ as defined in Eq.~(\ref{eq:residual}).
In this way, there is no need to make assumptions on the relations between the
inclusive charged hadron fragmentation functions -- necessary for fitting
purposes -- which may hold for one of the most abundant hadron species
like pions, but not necessarily for the others.
Factoring out the contributions from charged pion, kaon, and (anti-)proton
fragmentation functions, we are left with a comparatively small
residual contribution $D_i^{res^{\pm}}$, whose flavor symmetry assumptions, see Sec.~\ref{sec:chghad},
have very little impact on the inclusive charged hadron fragmentation functions
and the quality and reliability of the fit.

We begin by discussing the remarkable measurements of single-inclusive
charged hadron production in $e^+e^-$ annihilation,
collected by a variety of experiments
\cite{ref:tpcdata,ref:slddata,ref:alephdata,ref:delphidata,ref:tassodata,ref:opal}.
Figures~\ref{fig:hadroninclusive} and \ref{fig:hadron-tagged} show detailed comparisons between
the available data and the results of adding the already known pion, kaon, and proton
fragmentation functions to the outcome of the fit for the
residual hadron contributions $D_i^{res^{\pm}}$.

Tables \ref{tab:nlorespara} and \ref{tab:exprestab}, show the values
obtained for the parameters in Eq.~(\ref{eq:ff-input}) specifying the
optimum fit of residual charged hadron fragmentation functions $D_i^{res^+}(z,\mu_0)$ at
NLO accuracy and summarize the $\chi^2$ values for each
individual set of data included in the fit, respectively.
As in Tab.~\ref{tab:expprotontab}, for each set of data,
values of the normalization factors resulting from the fit
and constrained to vary within range quoted by experiment,
are also given in Tab.~\ref{tab:exprestab}.

All data sets shown in Figs.~\ref{fig:hadroninclusive} and \ref{fig:hadron-tagged}
are very well reproduced by our new fit, again closely following the
trend of the data again also below $z_{min}=0.1$.
This is most apparent in the (data-theory)/theory comparisons.
The impressive quality of the fit serves as an important cross-check of the consistency of entire
fitting procedure for all hadron species since the main contributions to the inclusive charged hadron
cross sections in Figs.~\ref{fig:hadroninclusive} and \ref{fig:hadron-tagged}
are already fixed by the pion, kaon, and proton fits. For instance, overshooting
the data by just summing up pion, kaons, and protons would have pointed to some serious
inconsistency in the global analysis. As expected, we also find that the residual contribution
from $D_i^{res^{\pm}}$ is indeed small and -- another non-trivial check -- positive,
see below.
Note that we estimate an average uncertainty of $5\%$
in all theoretical calculations of the inclusive charged hadron cross sections
stemming from the propagation of uncertainties of pion, kaon, and proton fragmentation functions.
This additional uncertainty is included in the $\chi^2$ minimization procedure
for determining $D_i^{res^{\pm}}$.
%
\begin{table}[thb!]
\caption{\label{tab:nlorespara}Parameters describing the NLO
fragmentation functions for positively charged
residual hadrons, $D_i^{res+}(z,\mu_0)$, at the input scale
$\mu_0=1\,\mathrm{GeV}$.
Inputs for the charm and bottom fragmentation functions refer to
$\mu_0=m_c=1.43\,\mathrm{GeV}$ and $\mu_0=m_b=4.3\,\mathrm{GeV}$, respectively.}
\begin{ruledtabular}
\begin{tabular}{cccccc}
flavor $i$ &$N_i$ & $\alpha_i$ & $\beta_i$ &$\gamma_i$ &$\delta_i$\\
\hline
$u+\overline{u}$ & 0.0038&10.000& 1.20& 0.0003&18.51\\
$\overline{u}$   & 0.0001&10.000&21.20& 0.0003&18.51\\
$c+\overline{c}$ & 0.0752& 0.406& 3.91& 0.0000& 0.00\\
$b+\overline{b}$ & 0.0936&-0.150& 3.61& 0.0000& 0.00\\
$g$              & 0.0001&-0.497& 9.99&20.000&14.75\\
\end{tabular}
\end{ruledtabular}
\caption{\label{tab:exprestab}Data used in the NLO global analysis of
residual charged hadron fragmentation functions, the individual $\chi^2$ values for
each set, the fitted normalizations, and the total $\chi^2$ of the fit.}
\begin{ruledtabular}
\begin{tabular}{lcccc}
experiment& data & rel.\ norm.\ &data points & $\chi^2$ \\
          & type & in fit       & fitted     &         \\\hline
TPC \cite{ref:tpcdata}  & incl.\  &  1.027 & 17 & 1.7 \\
SLD \cite{ref:slddata}  & incl.\  &  1.006 & 21 & 13.2 \\
ALEPH \cite{ref:alephdata}& incl.\ & 1.027 & 27 & 27.0 \\
DELPHI \cite{ref:delphidata}  & incl.\  & 1.0  & 12 & 6.8 \\
          & ``$uds$ tag''   &  1.0  & 12 & 7.6 \\
          & ``$b$ tag''     &  1.0  & 12 & 4.9 \\
TASSO \cite{ref:tassodata}  & incl.\ (44 GeV)  & 1.0  & 14 & 11.5 \\
                            & incl.\ (35 GeV)  & 1.0  & 14 & 19.9 \\
OPAL \cite{ref:opal}   & incl.\  & 1.0  & 12 & 5.7 \\
          & ``$uds$ tag'' &  1.0  & 12 & 11.9\\
          & ``$c$ tag'' &  1.0  & 12 & 7.4 \\
          & ``$b$ tag'' &  1.0  & 12 & 3.9 \\ \hline
ALEPH \cite{ref:alephdata}& long.\ incl.\ & 1.0 & 11 & 1.8 \\
OPAL  \cite{ref:opall}   & long.\ incl.\  & 1.0  & 12 & 3.3 \\
DELPHI \cite{ref:delphil}  & long.\ incl.\  & 1.0  & 12 & 11.6 \\
                          & long.\ ``$uds$ tag'' & 1.0  & 12 & 35.1\\
                           & long.\ ``$b$ tag'' &  1.0  & 12 & 4.7 \\\hline
EMC \cite{ref:emc}  & $h^+$    &  0.987 & 98 & 99.1 \\
                               & $h^-$ &  0.987 & 99 & 156.8 \\\hline
CDF \cite{ref:tev}  & 630 GeV & 1.1  & 16  &  103.4             \\
                         & 1.8 TeV & 1.1  & 37  &  112.7             \\
UA1 \cite{ref:ua1}  &   200 GeV &  1.1 & 31 &  111.5              \\
                       &   500 GeV & 1.1 & 32 &  44.5               \\
                       &   630 GeV  & 1.1 & 41 &  214.2            \\
                       &   900 GeV & 1.1 & 44   &  118.1            \\
UA2 \cite{ref:ua2}  &    540 GeV & 1.1 & 27 &  89.3               \\
 \hline\hline
{\bf TOTAL:} & & & 661  & 1227.6 \\
\end{tabular}
\end{ruledtabular}
\end{table}

\begin{figure*}[th!]
\begin{center}
\vspace*{-0.6cm}
\epsfig{figure=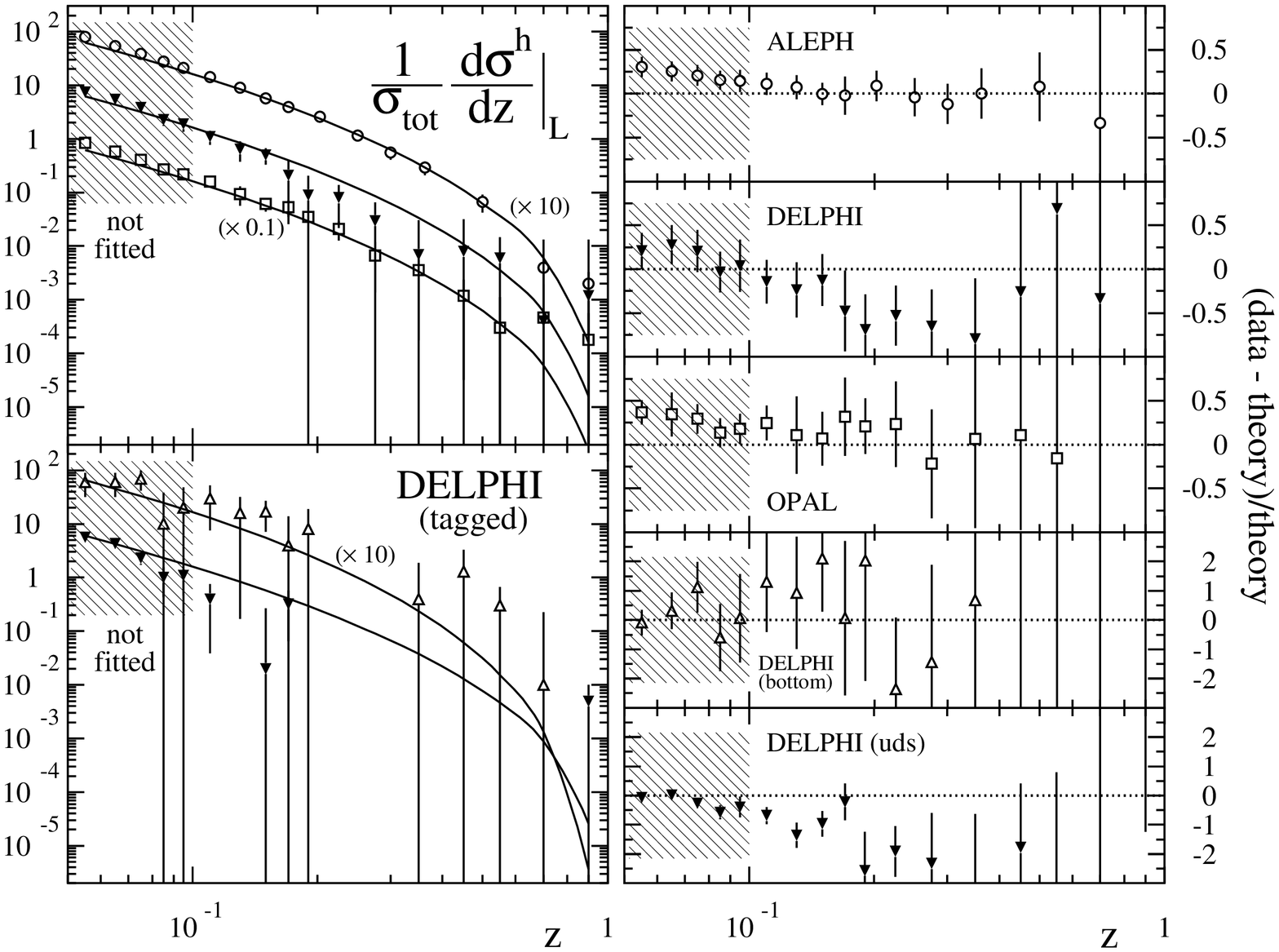,width=0.7\textwidth}
\end{center}
\vspace*{-0.7cm}
\caption{Same as in Figs.~\ref{fig:hadroninclusive} and \ref{fig:hadron-tagged} but now for the
longitudinal cross sections.
\label{fig:hadron-long}}
\begin{center}
\vspace*{-0.6cm}
\epsfig{figure=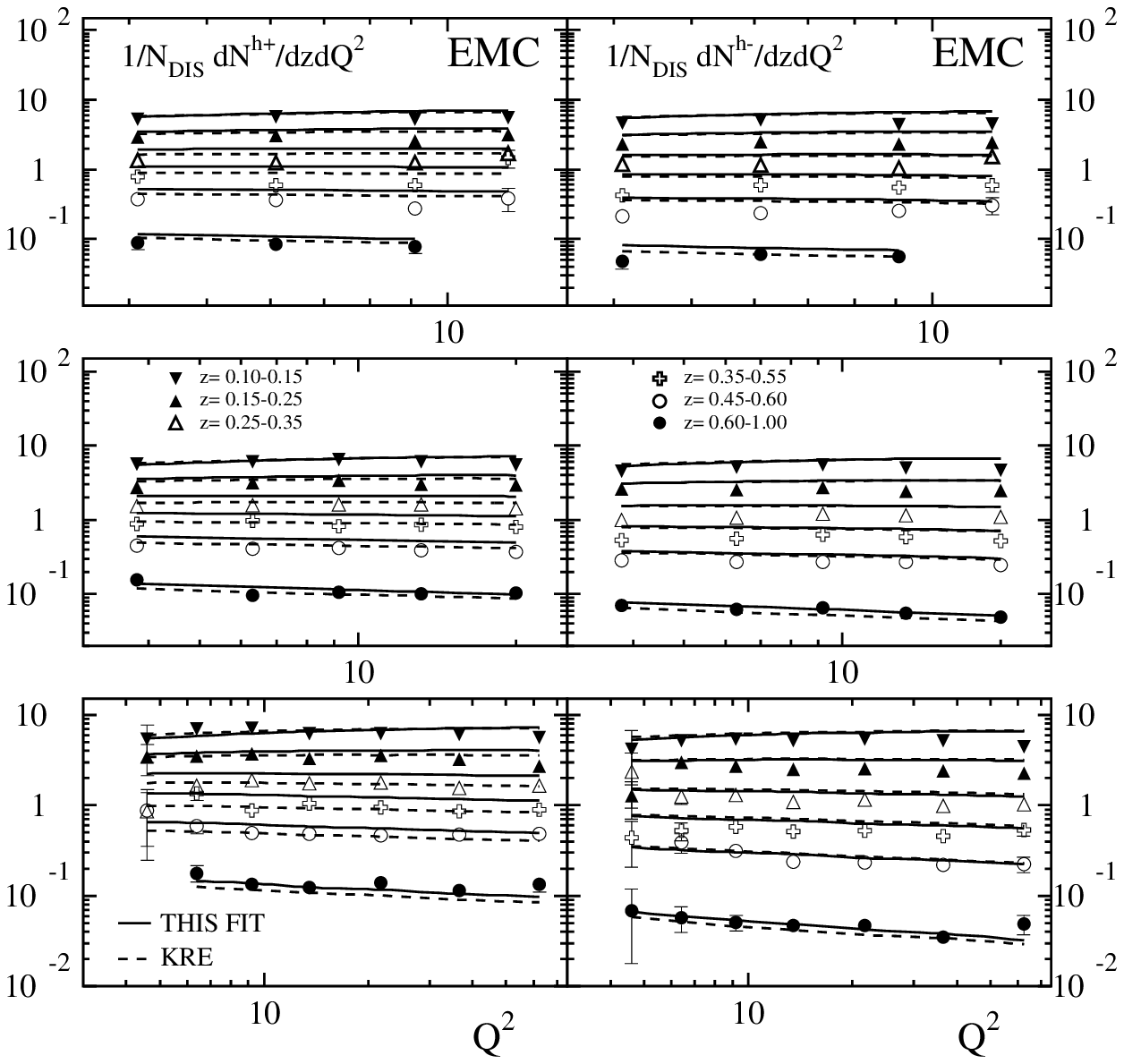,width=0.75\textwidth}
\end{center}
\vspace*{-0.7cm}
\caption{Comparison of our NLO results (solid lines) for the positively and negatively charged hadron
multiplicities in SIDIS, $(1/N_{\mathrm{DIS}}) dN^{h^{\pm}}/dz dQ^2$, with data
from EMC \cite{ref:emc} for different bins in $z$. Upper, middle and lower panels correspond to
beam energies of 120, 200, and 280 GeV, respectively.
Also shown are the results obtained with the KRE \cite{ref:kretzer} parameterization
(dashed lines).
\label{fig:sidish}}
\vspace*{-0.5cm}
\end{figure*}
As can be inferred from Tab.~\ref{tab:exprestab}, our global analysis contains
also several other data sets which help to further constrain certain aspects
of the residual charged hadron fragmentation functions and, more importantly,
provide additional consistency checks.
\begin{figure*}[thb!]
\begin{center}
\vspace*{-0.6cm}
\epsfig{figure=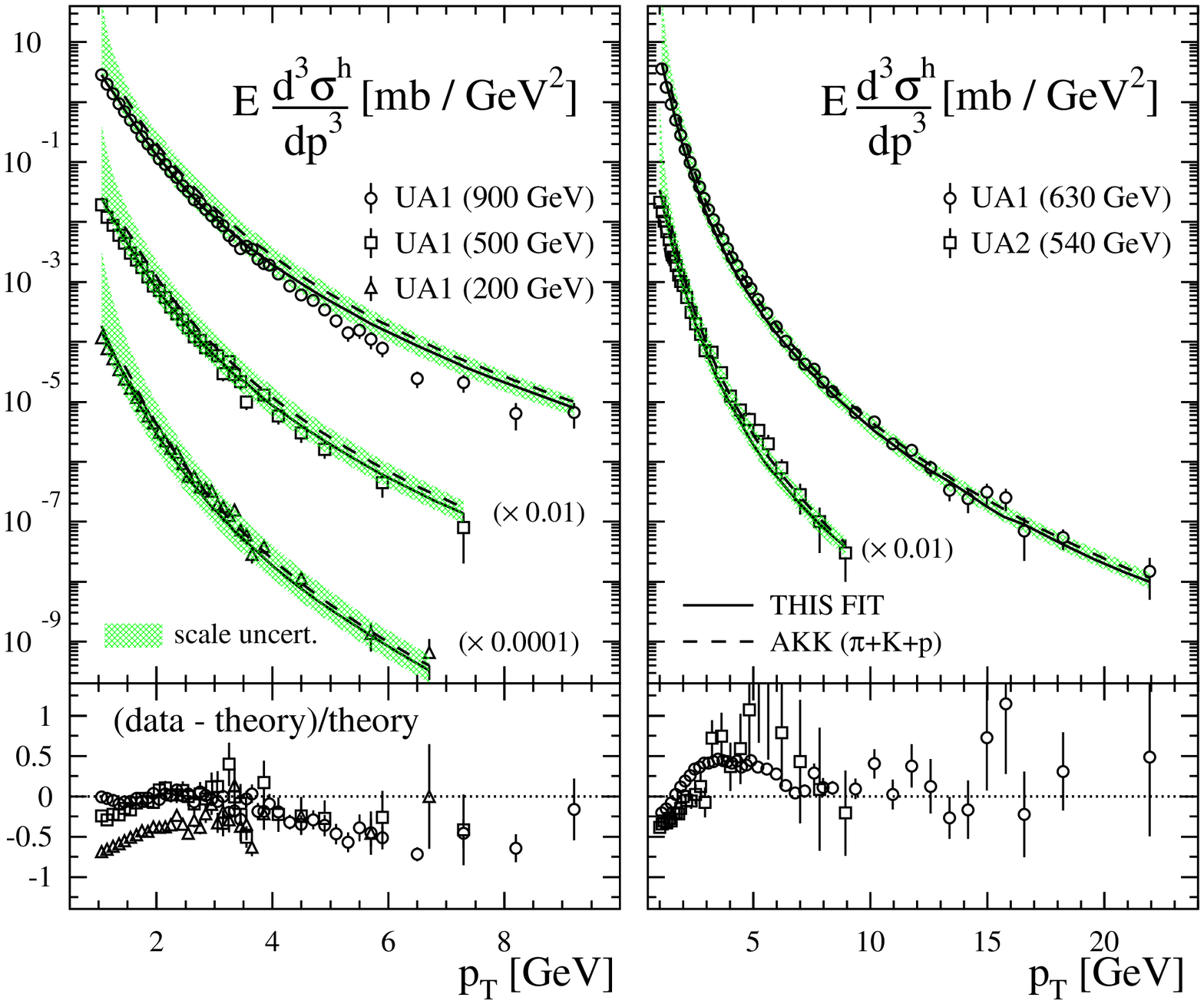,width=0.7\textwidth}
\end{center}
\vspace*{-0.7cm}
\caption{Upper panels: comparison of our NLO results for
single-inclusive charged hadron production
$p\bar{p}\rightarrow hX$, where $h=(h^+ + h^-)/2$,
with data from UA1 \cite{ref:ua1} and UA2 \cite{ref:ua2} for various
c.m.s.\ energies $\sqrt{s}$ using $\mu_f=\mu_r=p_T$.
The UA1 data cover $|y|<2.5$, except for $630\,\mathrm{GeV}$, where
$|y|<3.0$ and also $h=(h^+ + h^-)$. For UA2 the rapidity range
$1.0\le |y|\le 1.8$ is covered.
Also shown are the results obtained with the AKK \cite{ref:akk} parametrization
including only pions, kaons, and (anti-)protons.
The shaded bands indicate theoretical uncertainties when all
scales are varied in the range $p_T/2\le \mu_f=\mu_r \le 2 p_T$.
Lower panels: ``(data-theory)/theory'' for our NLO results.
\label{fig:cern-proton}}
\vspace*{-0.5cm}
\end{figure*}
Longitudinal cross section measurements in $e^+e^-$ annihilation are 
expected to help to constrain gluon fragmentation since, contrary to 
inclusive total
cross sections, a non-vanishing ${\cal{O}}(\alpha_s)$ gluon coefficient function is already
present at LO. For our NLO fits of the longitudinal cross section, we
include the ${\cal{O}}(\alpha_s^2)$ coefficient functions \cite{ref:fl}.
In Fig. \ref{fig:hadron-long} we compare the result from the NLO global analysis
for the longitudinal inclusive and flavor tagged cross section with the available
LEP data \cite{ref:alephdata,ref:opall,ref:delphil}. With the exception of the ALEPH
data, the statistical precision is not really impressive, in particular, in the
fitted region $z>z_{\min}$, but all data are reasonably well described by the fit.

Data on charged pion and kaon multiplicities in SIDIS have provided valuable
information on scale dependence and the charge and flavor separation of the corresponding fragmentation
functions in our previous analysis \cite{deFlorian:2007aj}.
For inclusive charged hadrons, the data collected by EMC \cite{ref:emc} cover
a rather large kinematic region with very narrow $Q^2$-bins, what makes it particularly 
suitable for a QCD fit. As expected, these data prove to be very valuable
in our global analysis here.
Figure~\ref{fig:sidish} shows the outcome of the fits compared to the EMC multiplicities
for positively and negatively charged hadrons covering all beam energies of 120, 200,
and 280 GeV. The good precision of the data over the entire $z$ range helps to
further constrain the high $z$ behavior of the quark and anti-quark fragmentation functions.
An interesting thing to notice is that even though the KRE sets for pions and kaons
overestimate the corresponding SIDIS multiplicities as demonstrated in \cite{deFlorian:2007aj}, this
effect is somehow compensated for in the charged hadron multiplicities shown in
Fig.~\ref{fig:sidish} at the expense of comparatively small residual fragmentation functions.
In order to account for the error introduced by the finite size
of the bins in $z$ and $Q^2$,  as well as the uncertainties introduced by the use of the
pion, kaon, and proton fragmentation functions, we include an additional $10\%$ theoretical
uncertainty in the $\chi^2$ minimization. There is also an additional $11\%$
experimental normalization uncertainty \cite{ref:emc} not included in the errors bars shown in Fig.~\ref{fig:sidish}.

\begin{figure}[thb!]
\begin{center}
\vspace*{-0.6cm}
\epsfig{figure=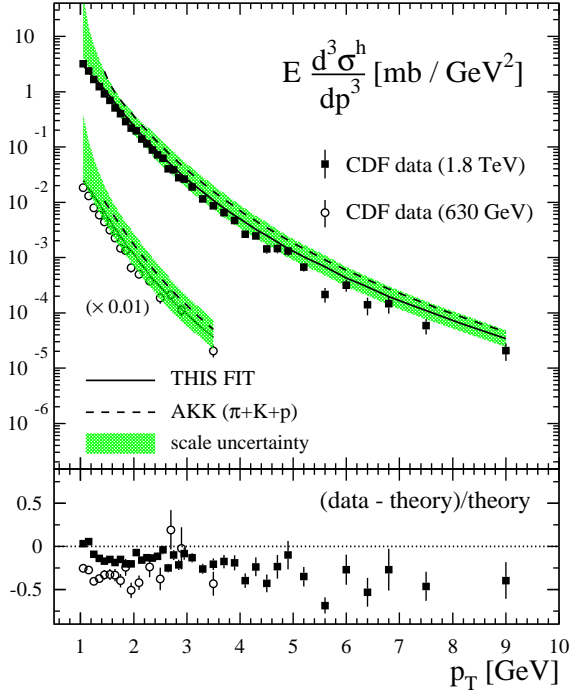,width=0.49\textwidth}
\end{center}
\vspace*{-0.7cm}
\caption{Similar to Fig.~\ref{fig:cern-proton}
but now comparing our NLO results to the CDF data at two different
c.m.s.\ energies $\sqrt{s}$. The data are for $h=(h^+ + h^-)/2$ and cover
$|y|\le 1$ in rapidity.
\label{fig:tev-proton}}
\vspace*{-0.5cm}
\end{figure}
The final ingredient to our global analysis of fragmentation functions $D_i^{h^{\pm}}$
for unidentified charged hadrons are single-inclusive hadroproduction data from $p\bar{p}$ collisions
measured by the UA1 \cite{ref:ua1} and UA2 \cite{ref:ua2} collaborations at CERN, 
and by the CDF collaboration \cite{ref:tev} at Fermilab's TeVatron. We have not included 
proton-proton collision data from fixed target experiments, since the validity of fixed order 
calculations 
in pQCD at lower energies is seriously in doubt \cite{de Florian:2005yj}.
The data span a range of c.m.s.\ energies $\sqrt{s}$ from 200~GeV to 1.8~TeV but
do not discriminate different hadron charges. Due to the dominance of gluon-induced
processes at the available small-to-medium values of the hadron's $p_T$, they mainly
probe the gluon fragmentation function $D_g^{h^{\pm}}$.
To give an example, in the case of UA1 at $\sqrt{s}=630\,\mathrm{GeV}$, where
the available data cover the largest range of transverse momenta $p_T$ from 1 GeV to 22 GeV,
in which the gluonic contribution decreases from around 90$\%$ in the lowest $p_T$ bin 
to a sizable 40$\%$ at the highest transverse momentum.
Figures \ref{fig:cern-proton} and \ref{fig:tev-proton} show the comparisons between
the results of our NLO fits to the single-inclusive cross sections and the experimental
data from CERN and Fermilab, respectively.
Within the fairly large theoretical scale ambiguities, indicated by the shaded bands,
which refer to varying the factorization and renormalization scales simultaneously
in the range $p_T/2\lesssim \mu_f=\mu_r\lesssim 2 p_T$, the overall agreement between
theory and data is reassuringly good.
The obtained gluon fragmentation functions into charged pions, kaons, and (anti-)protons,
which dominate the single-inclusive cross sections shown in
Figs.~\ref{fig:cern-proton} and \ref{fig:tev-proton}, are consistent with data, and
the residual contribution $D_g^{res^{\pm}}$ is very small.
Again, we include a $5\%$ theoretical error due to the choice of the scale in
the $\chi^2$ minimization and another $5\%$
associated with the propagation of the uncertainties of pion, kaon, and proton fragmentation functions.

\begin{figure}[th!]
\begin{center}
\vspace*{-0.6cm}
\epsfig{figure=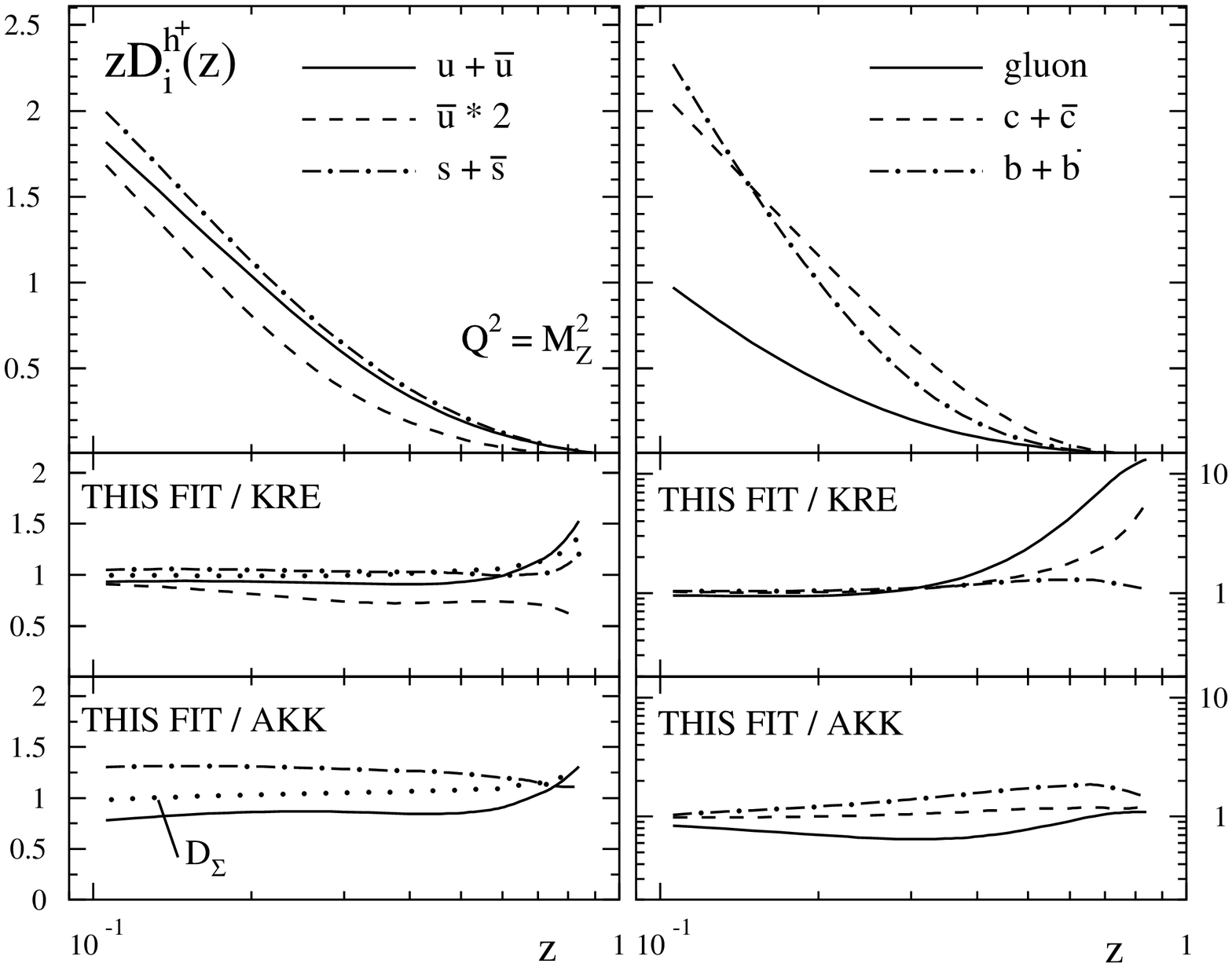,width=0.5\textwidth}
\end{center}
\vspace*{-0.7cm}
\caption{Upper panels: individual fragmentation functions for positively
charged hadrons $zD_i^{h^+}(z,Q^2)$ at $Q^2=M_Z^2$
for $i=u+\bar{u},\, 2\bar{u},\, s+\bar{s},\, g,\, c + \bar{c}$, and
$b+\bar{b}$.
Middle panels: ratios of our fragmentation functions to the ones
of KRE \cite{ref:kretzer}. The dotted line indicates the ratio for singlet
combination of fragmentation functions $zD_{\Sigma}^{h^{+}}$.
Lower panels: ratios of our fragmentation functions to the ones
of AKK \cite{ref:akk}; note that $D_{\bar{u}}^{h^{+}}$ is not
available in the AKK analysis.
\label{fig:ff-hadron}}
\begin{center}
\vspace*{-0.6cm}
\epsfig{figure=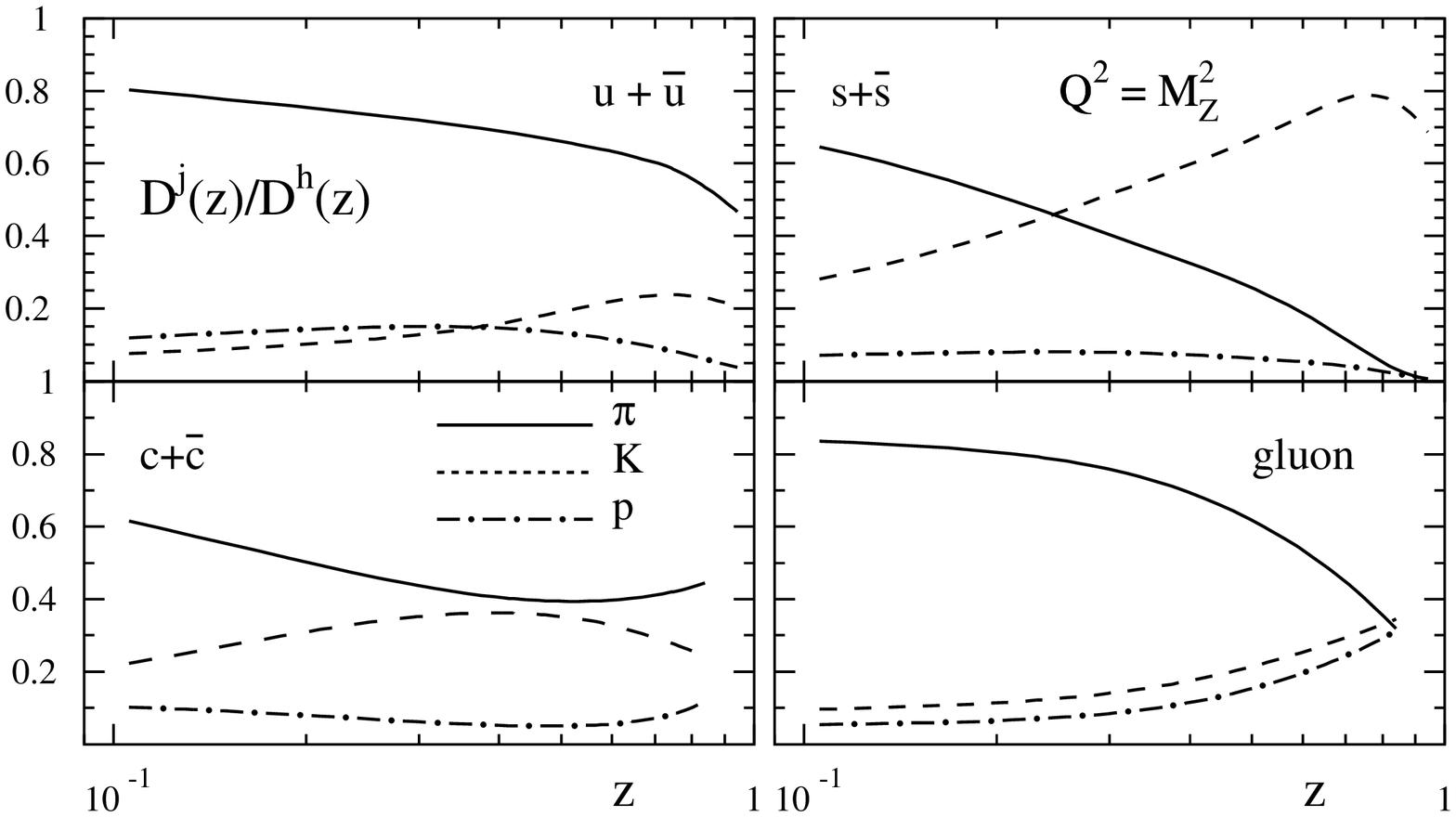,width=0.5\textwidth}
\end{center}
\vspace*{-2.7cm}
\caption{Partial contributions of pions, kaons, and protons
to the charged hadron fragmentation functions as a function of $z$ at $Q^2=M_Z^2$.
\label{fig:ff-comp}}
\vspace*{-0.5cm}
\end{figure}

The resulting fragmentation functions for positively charged hadrons $D_i^{h^{+}}$,
evolved to the scale $M_Z$, are shown in Fig.~\ref{fig:ff-hadron} and compared to the
distributions of KRE \cite{ref:kretzer} and AKK \cite{ref:akk}. Recall that in the
AKK analysis $D_i^{h^{+}}$ is approximated by the sum of pion, kaon, and
proton fragmentation functions. Compared to sets of KRE and AKK, we find again good agreement
in the total singlet contribution $D_{\Sigma}^{h^{+}}$, but some noticeable differences in
the charge and flavor separation, for instance, in $D_{\bar{u}}^{h^{+}}$,
and in the gluon fragmentation $D_g^{h^{+}}$.
In general, the differences become larger towards $z\to 1$, which has been already
observed for the pion, kaon \cite{deFlorian:2007aj}, and proton fragmentation functions
(Fig.~\ref{fig:ff-proton}).

One striking feature of the distributions shown in Fig.~\ref{fig:ff-hadron}
is the only slightly broken flavor democracy for $q+\bar{q}$ fragmentation functions
into charged hadrons.
To understand this better, the partial contributions of pions, kaons, and protons
to the charged hadron fragmentation functions are shown in Fig.~\ref{fig:ff-comp}.
As can be noticed, the $u$ quark and gluon fragmentation functions are
completely dominated by the pion contribution, except for
very large momentum fractions $z$, where proton-proton scattering data require some
increase in the kaon and proton part. The strange quark fragmentation
function for charged hadrons in the valence region is, as expected, dominated by kaons,
while both kaons and pions contribute with a similar amount to the charm (and bottom)
distributions. The residual contribution becomes sizable only for the heavy quark 
fragmentation functions.

\begin{figure}[thb!]
\begin{center}
\vspace*{-.6cm}
\epsfig{figure=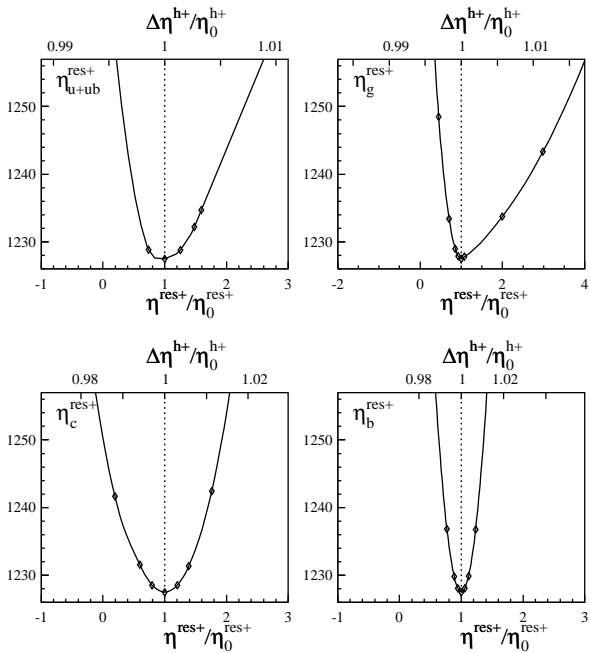,width=0.55\textwidth}
\end{center}
\vspace*{-.7cm}
\caption{Profiles of $\chi^2$ for the NLO residual charged hadron
fragmentation fit as a function of the truncated second moments
$\eta^{res^+}_i(x_p=0.2,\,Q^2=50\,\text{GeV}^2)$ for different flavors.
The moments are normalized to the value $\eta^{res^+}_{i\,0}$ they
take in the best fit to data. The upper horizontal scales show
variations relative to the change in the moments of the total
charged hadron fragmentation functions $\eta^{h^{+}}$.
\label{fig:hadron-uncert}}
\vspace*{-0.5cm}
\end{figure}
Even though the relative uncertainties of the residual charged hadron
fragmentation functions $D_i^{res^{\pm}}$ are rather large, exceeding even
$200\%$ within a conservative $2\%$ increase $\Delta\chi^2$
in the total $\chi^2$ of the fit, the effect on $D_i^{h+}$ is fairly small
and amounts to less than about $2\%$. This is readily explained by the relative
smallness of the residual charged hadron contribution to the sum in
Eq.~(\ref{eq:residual}).
The best constrained residual densities are those for charm and bottom,
which are also the most significant ones in absolute size,
although almost innocuous for SIDIS and $p\bar{p}$ data.
In Fig.~\ref{fig:hadron-uncert} we show the $\chi^2$-profiles of the global fit
as a function of the range of variation of the truncated second moments of the
residual charged hadron fragmentation functions $D_i^{res^{+}}$, see Eq.~(\ref{eq:truncmom}).
In the upper horizontal scales we also show the impact of those variations
relative to the change in the moments of the total
charged hadron fragmentation functions $D_i^{h^{+}}$.
\begin{table}[thb!]
\caption{\label{tab:loprotonpara} As in Tab.~\ref{tab:nloprotonpara}
but now describing the LO fragmentation functions into protons,
$D_i^{p}(z,\mu_0)$, at the input scale.}
\begin{ruledtabular}
\begin{tabular}{cccccc}
flavor $i$ &$N_i$ & $\alpha_i$ & $\beta_i$ &$\gamma_i$ &$\delta_i$\\
\hline
$u+\overline{u}$& 0.094& 0.041& 1.485&15.0& 3.44\\
$d+\overline{d}$& 0.059& 0.041& 1.485&15.0& 3.44\\
$\overline{u}$&  0.036& 0.041& 1.998&15.0& 3.44\\
$\overline{d}$&  0.022& 0.041& 1.998&15.0& 3.44\\
$s+\overline{s}$&  0.045& 0.041& 1.998&15.0& 3.44\\
$c+\overline{c}$& 0.079&-0.887& 5.436& 0.0& 0.00\\
$b+\overline{b}$& 0.047& 0.103&10.00& 0.0& 0.00\\
$g$& 0.029& 6.000& 1.200& 0.0& 0.000\\
\end{tabular}
\end{ruledtabular}
%
\vspace*{0.4cm}
\caption{\label{tab:lorespara} As in Tab.~\ref{tab:nlorespara} but now for
the LO fragmentation functions into positively charged residual hadrons
$D_i^{res^+}$.}
\begin{ruledtabular}
\begin{tabular}{cccccc}
flavor $i$ &$N_i$ & $\alpha_i$ & $\beta_i$ &$\gamma_i$ &$\delta_i$\\
\hline
$u+\overline{u}$& 0.0002& 9.986& 1.543&15.0&19.90\\
$\overline{u}$  & 0.0001& 9.986&21.543&15.0&19.90\\
$c+\overline{c}$& 0.147&-0.051& 2.792& 0.0& 0.00\\
$b+\overline{b}$& 0.113&-0.574& 2.949& 0.0& 0.00\\
$g$& 0.0001&-0.499&10.000&20.0&19.64\\
\end{tabular}
\end{ruledtabular}
\vspace*{0.4cm}
\caption{\label{tab:expprotonlotab} Same as in Tab.~\ref{tab:expprotontab}
but now at LO accuracy.}
\begin{ruledtabular}
\begin{tabular}{lcccc}
experiment& data & rel.\ norm.\ &data points & $\chi^2$ \\
          & type & in fit       & fitted     &         \\\hline
TPC \cite{ref:tpcdata}  & incl.\  &  1.043  & 12 & 7.5 \\
SLD \cite{ref:slddata}  & incl.\  &  0.983 & 18 & 11.8 \\
          & ``$uds$ tag''         &  0.983 & 9  & 10.7 \\
          & ``$c$ tag''           &  0.983 & 9  &  9.6  \\
          & ``$b$ tag''           &  0.983 & 9  &  9.3 \\
ALEPH \cite{ref:alephdata}    & incl.\  & 0.97 & 13 &  11.6 \\
DELPHI \cite{ref:delphidata}  & incl.\  & 1.0  & 12 & 3.9 \\
          & ``$uds$ tag''   &  1.0  & 12 & 0.7 \\
          & ``$b$ tag''     &  1.0  & 12 & 9.0 \\
OPAL \cite{ref:opaleta}
          & ``$u$ tag'' &  1.10  & 5 & 7.8 \\
          & ``$d$ tag'' &  1.10  & 5 & 12.8  \\
          & ``$s$ tag'' &  1.10  & 5 & 5.4 \\
          & ``$c$ tag'' &  1.10  & 5 & 5.0 \\
          & ``$b$ tag'' &  1.10  & 5 & 5.7 \\\hline
STAR \cite{ref:starproton}  & $p$  & 0.95  & 14  & 42.9           \\
                          & $\bar{p}$ & 0.95  & 14  & 32.4           \\
 \hline\hline
{\bf TOTAL:} & & & 159 & 186.1\\
\end{tabular}
\end{ruledtabular}
\end{table}

\section{Conclusions}
By extending our previous global analyses for pions and kaons to the case
of (anti-)protons and unidentified charged hadrons, we have completed a
comprehensive study of single-inclusive hadron production within
pQCD at NLO accuracy.
%
\begin{table}[th!]
\caption{\label{tab:loexprestab} Same as in Tab.~\ref{tab:exprestab}
but now at LO accuracy.}
\begin{ruledtabular}
\begin{tabular}{lcccc}
experiment& data & rel.\ norm.\ &data points & $\chi^2$ \\
          & type & in fit       & fitted     &         \\\hline
TPC \cite{ref:tpcdata}  & incl.\  &  1.041 & 17 & 7.7 \\
SLD \cite{ref:slddata}  & incl.\  &  1.014 & 21 & 63.0 \\
ALEPH \cite{ref:alephdata}& incl.\ & 1.013 & 27 & 46.4 \\
DELPHI \cite{ref:delphidata}  & incl.\  & 1.0  & 12 & 12.2 \\
          & ``$uds$ tag''   &  1.0  & 12 & 22.3 \\
          & ``$b$ tag''     &  1.0  & 12 & 16.6 \\
TASSO \cite{ref:tassodata}  & incl.\ (44 GeV) & 1.0  & 14 & 22.3 \\
                            & incl.\ (35 GeV) & 1.0  & 14 & 56.3 \\
OPAL \cite{ref:opal}   & incl.\  & 1.0  & 12 & 28.5 \\
          & ``$uds$ tag'' &  1.0  & 12 & 32.9\\
          & ``$c$ tag'' &  1.0  & 12 & 18.1 \\
          & ``$b$ tag'' &  1.0  & 12 & 8.6 \\ \hline
ALEPH  \cite{ref:alephdata}& long.\ incl.\ & 1.0 & 11 & 11.5 \\
OPAL   \cite{ref:opall}   & long.\ incl.\  & 1.0  & 12 & 8.1 \\
DELPHI  \cite{ref:delphil}  & long.\ incl.\  & 1.0  & 12 & 7.2 \\
                          & long.\ ``$uds$ tag'' & 1.0  & 12 & 28.0\\
                           & long.\ ``$b$ tag'' &  1.0  & 12 & 5.0 \\\hline
EMC \cite{ref:emc}  & $h^+$    &  1.0 & 98 & 183.0 \\
                               & $h^-$ &  1.0 & 99 & 289.1 \\\hline
CDF \cite{ref:tev}  & 630 GeV  & 1.1  & 16  &  222.4             \\
                         & 1.8 TeV  & 1.1  & 37 &  515.5             \\
UA1 \cite{ref:ua1}  &   200 GeV & 0.988 & 31 &  84.5              \\
                       &   500 GeV & 0.988 & 32 &  129.4               \\
                       &   630 GeV & 0.988 & 41 &  48.3            \\
                       &   900 GeV & 0.988 & 44   &  391.7            \\
UA2 \cite{ref:ua2}  &    540 GeV & 0.988 & 27 &  135.3               \\
\hline\hline
{\bf TOTAL:} & & & 661  & 2393.9 \\
\end{tabular}
\end{ruledtabular}
\end{table}
Specifically, we have demonstrated that the pQCD framework based on
factorized cross sections can consistently account
for a large variety of processes with hadrons in the final-state
with remarkable precision, producing at the same time accurate
and {\em universal} sets of fragmentation functions.
The availability of these crucial non-perturbative inputs now
opens a wide range of possibilities for detailed studies
of the nucleon structure at NLO accuracy through single-inclusive
hadroproduction processes, encompassing the study of the spin and
flavor structure with polarized SIDIS measurements \cite{ref:polpdf},
in proton-proton collisions \cite{ref:jager1},
and in photoproduction \cite{ref:jager2}, the modification of parton
densities by nuclear effects \cite{deFlorian:2003qf}, or studies of
high $p_T$ electroproduction processes \cite{Daleo:2004pn}.

\acknowledgments
We warmly acknowledge Werner Vogelsang for helpful discussions,
and Carlos Garc\'{\i}a Canal for comments and suggestions.
This work was partially supported by CONICET, ANPCyT and UBACyT.

\section{Appendix: LO results}
%
For completeness, we have also performed global analyses
of the same sets of data given in Tables~\ref{tab:expprotontab}
and \ref{tab:exprestab}, where now all observables, $\alpha_s$, and
the scale evolution of the fragmentation functions are computed
at LO accuracy. We use the same parametrization (\ref{eq:ff-input}) and
fitting procedure as in the NLO case, outlined in Sec.~II. 
The parameters of the optimum LO fits for $D_i^{p}$ and $D_i^{res^+}$
are given in Tables \ref{tab:loprotonpara} and
\ref{tab:lorespara} while the $\chi^2$ values obtained for the
individual sets of data are compiled in
Tables~\ref{tab:expprotonlotab} and \ref{tab:loexprestab}, respectively.

It is important to notice that the total $\chi^2$ of the LO fits is significantly
worse than at NLO accuracy, in particular, in case of the inclusive
charged hadron data. Our LO sets should be used only for rough estimates of observables
where NLO corrections are not yet available, or in event generators based on matrix
elements at LO accuracy. Because of the limited usefulness of the LO sets, we refrain from
going into any further details here.



\begin{thebibliography}{99}
%
\bibitem{deFlorian:2007aj}
D.~de Florian, R.~Sassot and M.~Stratmann, Phys. Rev. {\bf D75}, 114010 (2007).
%
\bibitem{ref:tpcdata}  H.\ Aihara {\em et al.} (TPC Collaboration),
Phys. Rev. Lett. {\bf 61}, 1263 (1998); Phys. Lett. {\bf B184}, 299
(1987); X.-Q.\ Lu, Ph.D.\ thesis, Johns Hopkins University [Report No.\
UMI-87-07273, 1986].
%
\bibitem{ref:slddata} K.\ Abe {\em et al.} (SLD Collaboration),
Phys. Rev. {\bf D59}, 052001 (1999).
%
\bibitem{ref:alephdata} D.\ Buskulic {\em et al.} (ALEPH Collaboration),
Z. Phys. {\bf C66}, 355 (1995); Phys. Lett. {\bf B357}, 487 (1995);
R.\ Barate {\em et al.} (ALEPH Collaboration), Phys. Rept. {\bf 294}, 1 (1998).
%
\bibitem{ref:delphidata} P.\ Abreu {\em et al.} (DELPHI Collaboration),
Eur. Phys. J. {\bf C5}, 585 (1998).
%
\bibitem{ref:opaleta} G.\ Abbiendi {\em et al.} (OPAL Collaboration),
Eur. Phys. J. {\bf C16}, 407 (2000).
%
\bibitem{ref:kkp} B.A.\ Kniehl, G.\ Kramer, and B.\ P\"{o}tter,
Nucl. Phys. {\bf B582}, 514 (2000).
%
\bibitem{ref:akk} S.\ Albino, B.A.\ Kniehl, and G.\ Kramer,
Nucl. Phys. {\bf B725}, 181 (2005); {\bf B734}, 50 (2006).
%
\bibitem{ref:hirai} M.\ Hirai, S.\ Kumano, T.-H.\ Nagai, and
K.\ Sudoh, Phys. Rev. {\bf D75}, 094009 (2007).
%
\bibitem{ref:starproton} J.\ Adams {\em et al.} (STAR Collaboration),
Phys. Lett. {\bf B637}, 161 (2006).
%
\bibitem{ref:tassodata} W.\ Braunschweig {\em et al.} (TASSO Collaboration),
Z. Phys. {\bf C42}, 189 (1989).
%
\bibitem{ref:opal} K.\ Ackerstaff  {\em et al.} (OPAL Collaboration),
Eur. Phys. J. {\bf C7}, 369 (1999).
%
\bibitem{ref:opall} R.\ Akers {\em et al.} (OPAL Collaboration),
Z. Phys. {\bf C68}, 203 (1995).
%
\bibitem{ref:delphil} P.\ Abreu {\em et al.} (DELPHI Collaboration),
Eur. Phys. J. {\bf C6}, 19 (1999).
%
\bibitem{ref:tev} F.~Abe {\it et al.} (CDF Collaboration),
Phys. Rev. Lett. {\bf 61}, 1819 (1988).
%
\bibitem{ref:ua1} C.~Albajar {\it et al.} (UA1 Collaboration),
Nucl. Phys. {\bf B335}, 261 (1990);
G.~Bocquet {\it et al.} (UA1 Collaboration),  Phys.\ Lett.\  {\bf B366}, 434 (1996).
%
\bibitem{ref:ua2} M.~Banner {\it et al.} (UA2 Collaboration),
Z.\ Phys.\  {\bf C27}, 329 (1985).
%
\bibitem{ref:emc} J.\ Ashman {\em et al.} (EMC Collaboration),
Z. Phys. {\bf C52}, 361 (1991).
%
\bibitem{ref:kretzer} S.\ Kretzer, Phys. Rev. {\bf D62}, 054001 (2000).
%
\bibitem{Bourhis:2000gs}
L.~Bourhis, M.~Fontannaz, J.~Ph.~Guillet, and M.~Werlen,
Eur. Phys. J. {\bf C19}, 89 (2001).
%
\bibitem{ref:mellin} M.\ Stratmann and W.\ Vogelsang, Phys. Rev. {\bf D64},
114007 (2001).
%
\bibitem{ref:brahmsdata} I.\ Arsene {\em et al.} (BRAHMS Collaboration),
arXiv:hep-ex/0701041.
%
\bibitem{ref:cteq} D.\ Stump {\em et al.}, Phys. Rev. {\bf D65}, 014012 (2002).
%
\bibitem{ref:fl}  P.J.\ Rijken and W.L.\ van Neerven, Nucl. Phys.
{\bf B487}, 233 (1997); Phys. Lett. {\bf B386}, 422 (1996);
A.~Mitov and S.O.~Moch, Nucl.\ Phys.\  {\bf B751}, 18 (2006).
%
\bibitem{de Florian:2005yj} D.~de Florian and W.~Vogelsang,
Phys.\ Rev.\  {\bf D71}, 114004 (2005).
%
\bibitem{ref:polpdf} D.~de Florian, G.A.~Navarro, and R.~Sassot,
Phys.\ Rev.\  {\bf D71}, 094018 (2005).
%
\bibitem{ref:jager1} D.\ de Florian, Phys. Rev. {\bf D67}, 054004 (2003);
B.\ J\"{a}ger, A.\ Sch\"{a}fer, M.\ Stratmann,
and W.\ Vogelsang, Phys. Rev. {\bf D67}, 054005 (2003).
%
\bibitem{ref:jager2} B.~J\"{a}ger, M.~Stratmann, and W.~Vogelsang, Phys.\ Rev.\  {\bf D68}, 114018 (2003);
Eur. Phys. J. {\bf C44}, 533 (2005).
%
\bibitem{deFlorian:2003qf} D.~de Florian and R.~Sassot,
Phys.\ Rev.\  {\bf D69}, 074028 (2004).
%
\bibitem{Daleo:2004pn} A.~Daleo, D.~de Florian, and R.~Sassot,
Phys.\ Rev.\ {\bf D71}, 034013 (2005).
%
\end{thebibliography}
\end{document}